\documentclass[twocolumn,twoside,slac_two]{revtex4}
\usepackage{graphicx}
\usepackage{fancyhdr}
\usepackage{threeparttable}
\usepackage{natbib}
\usepackage{booktabs}
\usepackage{amssymb}
\usepackage{epstopdf}

\def\mathbi#1{\textbf{\em #1}}

\begin{document}

\title{A Multiwavelength Cross-Correlation Variability Study of Fermi-LAT Blazars}

\author{V. Pati\~no-\'Alvarez, A. Carrami\~nana, L. Carrasco, V. Chavushyan}

\affiliation{$^1$Instituto Nacional de Astrof\'isica, \'Optica y Electr\'onica, Luis Enrique Erro 1, Tonantzintla, Puebla, 72840, M\'exico}

\begin{abstract}
\begin{small}
We carried out a multiwavelength cross-correlation analysis of a sample of 16 blazars detected by Fermi/LAT. The purpose is to investigate if there exists correlations between the distinct bands we analyze in this work. We searched for cross-correlated delays between emissions in optical, near-infrared and $\gamma$-ray bands for 16 blazars detected by Fermi-LAT, using three methods previously discussed in the literature: the interpolated cross-correlation function, the discrete cross-correlation function and the Z-transformed discrete cross-correlation function. Our results confirm the expectation that in all our sample the four NIR bands vary simultaneously. For three objects of our sample (3C 273, Mrk 501, and PMN J0808-0751), no correlation is found between any of the bands available for this study. For the remaining thirteen, a correlation was found between the V band and the NIR bands, indicating that in most of them the V band and the NIR bands vary simultaneously. For 4 objects (3C 454.3, PKS 0235+164, PKS 1510-089, and PKS 2155-304) a delayed correlation between the $\gamma$-ray emission and the NIR+V bands was found.
\end{small}
\end{abstract}

\maketitle

\thispagestyle{fancy}

\section{INTRODUCTION}
\begin{large}

Blazars are a subclass of Active Galactic Nuclei (AGN) characterized by a prominent jet  whose angle relative to the line of sight is very small, within a few degrees \citep{Bl:78a}. In many cases, these jets can be very luminous at wavelengths from radio to $\gamma$-rays. Due to the proximity of the jet axis to the line of sight, the emission from the jet is relativistically beamed, and hence amplified by an order of magnitude or more for some blazars. The observed time scales, in some cases, are also shorter than those in the rest frame of the jet. Most of the observed emission from radio to optical (sometimes UV) is due to synchrotron radiation from the jet \citep{Mar:98a}. X-rays and $\gamma$-rays may be produced via Inverse Compton scattering by the same energetic electrons radiating synchrotron emission (the so-called leptonic models, e.g. \citealp{Bo:07a}); or may be due to synchrotron radiation by protons co-accelerated with the electrons in the jet, interactions of these highly relativistic protons with external radiation fields, or proton-induced particle cascades (hadronic models, e.g. \citealp{Mu:03a}).

Time variability at multiple wavelengths is a defining characteristic of these objects, and has been used to probe the location and physical processes related to the emission at very fine resolutions (e.g. \citealp{Ma:08a}). Most of the observed blazars have the peak of their emission in the $\gamma$-ray part of the spectrum. However, until recently, long-term and well-sampled $\gamma$-ray light curves of blazars were not available. This has changed with the near-continuous monitoring activity of the Large Area Telescope (LAT) instrument on board the Fermi Gamma-Ray Space Telescope, launched in 2008, which provides the opportunity to study the variable SEDs of a large sample of blazars with truly simultaneous multi-frequency data.

In this contribution we present the results of a multiwavelength variability study on a sample of blazars observed by Fermi/LAT.

\section{THE SAMPLE AND OBSERVATIONAL DATA}

A sample of 16 blazars detected by Fermi/LAT was analyzed: 9 Optically Violent Variable quasars (OVV) and 7 BL Lacertae type objects (BLL) (see Table~\ref{sample}). 

We have light curves in Near InfraRed (NIR) J, H, K and Ks bands; and optical (V band) monitoring of Fermi/LAT Blazars. The time period covered by the observations is from the late 2007 up to mid-2012. The NIR data are from the Guillermo Haro Astrophysical Observatory (OAGH from spanish) using the Cananea Near-Infrared Camera (CANICA) and from from the Yale Fermi/ SMARTS project \citep{SMA:12a}; while the optical data (V Band) are from four different sources: most of the data is from the Ground-based Observational Support of the Fermi Gamma-Ray Space Telescope at the University of Arizona \citep{Steward} using the SPOL CCD Imaging/Spectropolarimeter and from the Yale Fermi/SMARTS project. Extra data for 3C 454.3 was taken from \cite{Ra:11a}, the WEBT Project; and V band data for QSO B0133+476 was taken from the MISAO Project \citep{MISAO}, specifically photometry taken by Miguel Rodriguez Marco with the SRO50 AAVSONet Robotic Telescope.

\begin{table}[t]
\begin{center}
\caption{The Sample}
\begin{small}
\begin{tabular}{lccc}
\hline Object & NED Classification & Redshift & $\rm \bar{V}$ \\
\hline
3C 66A          & BLL  & 0.444  & 14.4   \\
BL Lac          & BLL  & 0.069  & 14.7   \\
Mrk 421         & BLL  & 0.030  & 13.0   \\
Mrk 501         & BLL  & 0.034  & 13.9   \\
PKS 0716+714    & BLL  & 0.300  & 13.6   \\
PKS 2155-304    & BLL  & 0.116  & 13.5   \\
W Comae         & BLL  & 0.102  & 15.2   \\
\hline
3C 273          & OVV    & 0.158  & 12.8   \\
3C 279          & OVV    & 0.536  & 16.5   \\
3C 345          & OVV    & 0.593  & 17.2   \\
3C 454.3        & OVV    & 0.859  & 15.6   \\
PKS 0235+164    & OVV    & 0.940  & 18.0   \\
PKS 1510-089    & OVV    & 0.360  & 16.5   \\
PKS 1633+382    & OVV    & 1.814  & 17.4   \\
PMN J0808-0751  & OVV    & 1.837  & R$\sim$16.7   \\
QSO B0133+476   & OVV    & 0.859  & 17.0   \\
\hline
\end{tabular}
\end{small}
\label{sample}
\end{center}
\end{table}

\begin{figure*}[t]
\centering
\includegraphics[width=0.8\textwidth]{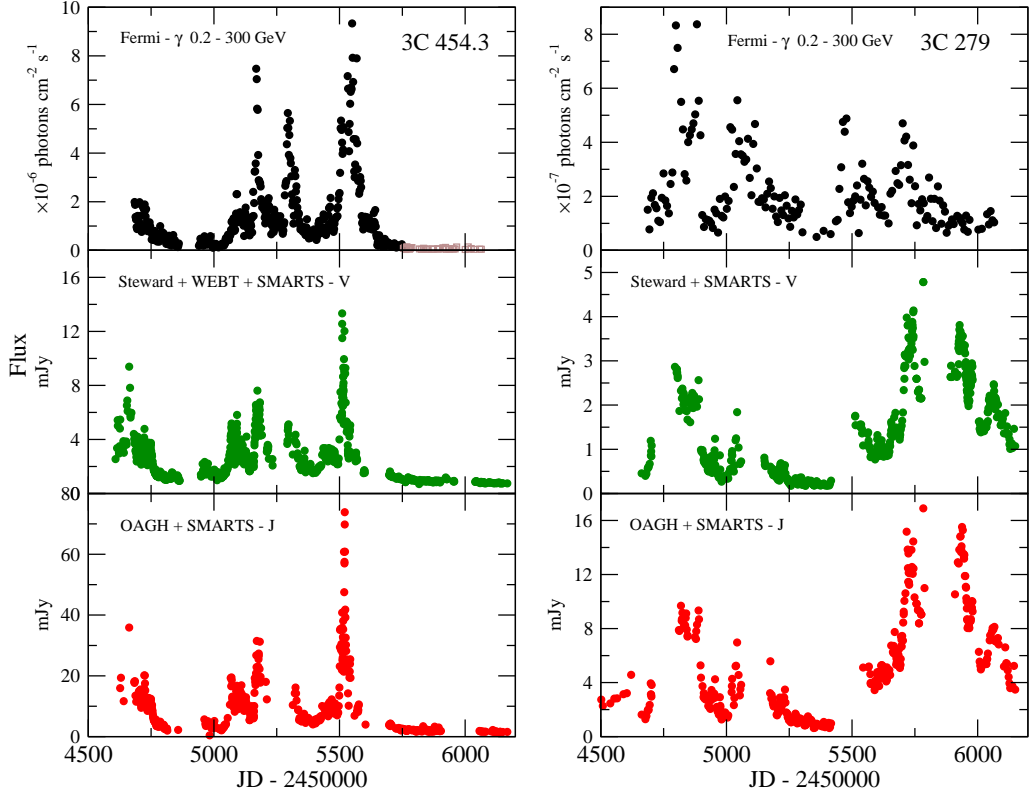}
\caption{Light Curves from 3C 454.3 and 3C 279 in $\gamma$-rays, V band and J band. Light brown squares in the 3C 454.3 $\gamma$-rays light curve are upper limits. We also have emission line and spectral continuum light curves from the Steward Observatory, which are being complemented with data from OAGH (see \citealp{PA:13b}).}
\label{Curvas}
\end{figure*}

\section{CROSS CORRELATION ANALYSIS}

A Cross-Correlation analysis of the light curves was carried out with the purpose of quantify lags between different kinds of emission, using three different methodologies of the cross correlation function: The Interpolated Cross Correlation Function (ICCF, \citealp{GS:86a}), The Discrete Cross Correlation Function (DCCF, \citealp{EK:88a}) and the Z-Transformed Discrete Cross Correlation Function (ZDCF, \citealp{TA:97a}).

\subsection{The Interpolated Cross Correlation Function}

The ICCF \citep{GS:86a} consists in interpolate the actual light curves, so to get the same number of points and equal sampling in both light curves. Once we have vectors of identical size, covering a time interval $T$ for both light curves, we must define a vector of lags which can run from $-T$ to $+T$ (in units of the spacing of the interpolated points); this vectors of lags represent movements that will have the time axis of one of the light curves. When we move the light curves in time, we compute the correlation coefficient only between the segments of the light curves that intersect; this is done to avoid problems for the non-stationarity of the light curves (see \citealp{WP:94a}). Note that if we displace the time axis of one of the light curves by a lag $+T$ or $-T$, we will be left with just one point in each light curve to compute the correlation coefficient, which is not statistically significant; therefore we make the vector of lags to run from $-0.8T$ to $+0.8T$ so we are left with at least 20\% of the light curve to compute a correlation coefficient.

Once we have our light curves $\mathbi{x}$ and $\mathbi{y}$, and a vector of lags $\mathbi{L}$, we can calculate the Cross-Correlation Function $P_{xy}(L)$ by using the Eq.~\ref{corrint} as defined by \cite{WF:76a}:

\begin{footnotesize}
\begin{equation}
P_{xy}(L) = \left\{ \begin{array}{ll}
\frac{\displaystyle
\sum_{k=0}^{N-|L|-1}(x_{k+|L|}-\bar{x})(y_k-\bar{y}) }
{\displaystyle
\sqrt{ \biggl[\sum_{k=0}^{N-1}(x_k-\bar{x})^2\biggr]
\biggl[\sum_{k=0}^{N-1}(y_k-\bar{y})^2\biggr] } } & For\ L<0\\

\frac{\displaystyle
\sum_{k=0}^{N-L-1}(x_k-\bar{x})(y_{k+L}-\bar{y}) }
{\displaystyle
\sqrt{ \biggl[\sum_{k=0}^{N-1}(x_k-\bar{x})^2\biggr]
\biggl[\sum_{k=0}^{N-1}(y_k-\bar{y})^2\biggr] } } & For\ L\ge0\\
\label{corrint}
\end{array} \right.
\end{equation}
\end{footnotesize}

where $\bar{x}$ is the mean of the points in the $\mathbi{x}$ curve that are involved in the calculation of the correlation coefficient, and similarly for $\bar{y}$. The means are recalculated for each lag.

\subsection{The Discrete Cross Correlation Function}

Unlike the ICCF, the method proposed by \cite{EK:88a} does not assume any understanding of the real light curve behavior; it uses only real data points which are separated by the time $\tau$. The correlation function itself is binned on time intervals $\delta t$, so that the value of the DCF at $\tau$ is the average over the interval $\tau-\delta t/2$ to $\tau+\delta t/2$.

If you have two vectors $\mathbi{a}$ and $\mathbi{b}$, then you can form pairs $(a_i,b_j)$, each one of these is associated with the pairwise lag $\Delta t_{ij}=t_j-t_i$. Now, for each of these pairs we can compute the unbinned discrete correlation functions:

\begin{equation}
UDCF_{ij}=\frac { (a_i-\bar{a})(b_j-\bar{b}) } { \sigma_a\sigma_b  }
\label{UDCF}
\end{equation}

where $\bar{a}$ and $\sigma_a$ are the means and standard deviations of the points $a_i$ in each bin, and in similar manner for series $\mathbi{b}$.

Originally \cite{EK:88a} used the entire series to calculate the means and the standard deviations, which is only applicable when we have stationary light curves, however, the AGN light curves are not stationary (i.e. the mean and standard deviation change with time, e.g. \citealp{WP:94a}). Also, they suggested that in order to preserve the proper normalization it is necessary to replace the $\sigma_a\sigma_b$ in the Eq.~\ref{UDCF} with $[(\sigma_a^2-e_a^2)(\sigma_b^2-e_b^2)]^{1/2}$. \cite{WP:94a}, however, argue that doing this greatly complicates any direct comparison with the interpolation results, which are unweighted, and therefore this makes an interpretation of the DCF amplitude less straightforward.

Next, if we have $M$ pairs for which $\tau-\delta t/2 \le \Delta t_{ij} < \tau+\delta t/2$, then the discrete correlation function of $\tau$ is:

\begin{equation}
DCF(\tau)=\frac{1}{M}\sum_{\tau-\delta t/2}^{\tau+\delta t/2}UDCF_{ij}(\Delta t_{ij})
\end{equation}

[Note that the $DCF(\tau)$ is not defined for a bin with no points]

Unlike the interpolation method, for which the errors are difficult to define, we can define almost directly a standard error for the DCF. If each of the individual UDCF$_{ij}$ within a single bin were totally uncorrelated, then the standard error in the determination of their mean would be:

\begin{small}
\begin{equation}
\sigma_{DCF}(\tau)=\frac{1}{\sqrt{M(M-1)}}\biggl\{\sum_{Bin}[UDCF_{ij}-DCF(\tau)]^2\biggr\}^{1/2}
\end{equation}
\end{small}

\subsection{The Z-Transformed Discrete Cross Correlation Function}

The Z-Transformed Discrete Correlation Function by \cite{TA:97a} is an alternative method for estimating the CCF of sparse, unevenly sampled light curves. The ZDCF corrects several biases of the discrete correlation function method of \cite{EK:88a} by using equal population binning and Fisher's z-transform (e.g. \citealp{KS:69a,KS:73a} and references therein).

If we have $n$ pairs in a given time-lag bin, the CCF($\tau$) is estimated by the correlation coefficient:

\begin{equation}
r=\frac{\sum_{i}^{n}(a_i-\bar{a})(b_i-\bar{b})/(n-1)}{s_as_b}
\end{equation}

where $\bar{a}$, $\bar{b}$ are the bin averages, and $s_a$, $s_b$ are the standard deviations of the points in a bin.

The sampling distribution of $r$ is highly skewed and far from normal, therefore estimating its sampling error by the simple variance $s_r$ can be very inaccurate.

If $\mathbi{a}$ and $\mathbi{b}$ are drawn from bivariate normal distributions, it is possible to transform $r$ into an approximately normally distributed random variable, Fisher's $z$. Defining:

\begin{small}
\begin{equation}
z=\frac{1}{2}\ln{\Bigg(\frac{1+r}{1-r}\Bigg)} \; \; , \; \; \;  \rho=\tanh{z} \; \; , \; \; \;  \zeta=\frac{1}{2}\ln{\Bigg(\frac{1+\rho}{1-\rho}\Bigg)}
\end{equation}
\end{small}

This yields to the mean of $z$ being approximately equal to:

\begin{small}
\begin{equation}
\bar{z}=\zeta+\frac{\rho}{2(n-1)}\times
\Bigg[ 1+\frac{5+\rho^2}{4(n-1)}+\frac{11+2\rho^2+3\rho^4}{8(n-1)^2} + \cdot\cdot\cdot\Bigg]
\end{equation}
\end{small}

and the variance of $z$ is approximately equal to:

\begin{small}
\begin{equation}
s_z^2=\frac{1}{n-1}\Bigg[ 1+\frac{4-\rho^2}{2(n-1)}+\frac{22-6\rho^2-3\rho^4}{6(n-1)^2}+\cdot\cdot\cdot \Bigg]
\end{equation}
\end{small}

 Transforming to $r$ again, the interval corresponding to the normal $\pm1\sigma$ error interval can be determined by:
 
 \begin{equation}
\delta r_{\pm}= |\tanh{(\bar{z}(r)\pm s_z(r))}-\rho|
\end{equation}

The binning method for the ZDCF is different from that of \cite{EK:88a}, in which the binning of the pairs is done in a fixed time interval, i. e. every bin has the same length in time, and the resulting DCF of that bin is the average of the UDCF of all pairs who fall into the bin; while, the ZDCF binning is for a fixed population, i. e. every bin has the same number of pairs, at least $n_{min}=11$ pairs, which is the minimum number for a meaningful statistical interpretation \citep{TA:97a}.

Also, in each bin, the interdependent pairs are discarded. This means that in a bin we can not have two pairs which use the same $a_i$ or $b_j$ element. Therefore, light curves with less than 12 points cannot be analyzed by this method.

We made a few modifications to the original method, by changing how the accommodation of pairs is done, we do not start from the median. Since many of the pairs found in the median of the $\Delta t_{ij}$ distribution use some of the same points of the light curves, a lot of pairs are discarded and not used in the actual analysis; therefore we start the accommodation of the pairs from the extreme negative part of the CCF (i.e. the pair with the smallest $\Delta t_{ij}$). Also, we require that all bins have the same number of pairs, unlike the original method where it was possible to obtain bins with different number of pairs; which may lead to different significance in each bin.

Examples of the CCF curves obtained with the three different statistical methods are shown in Fig.~\ref{CCF}

\begin{figure}[t]
\centering
\includegraphics[width=0.45\textwidth]{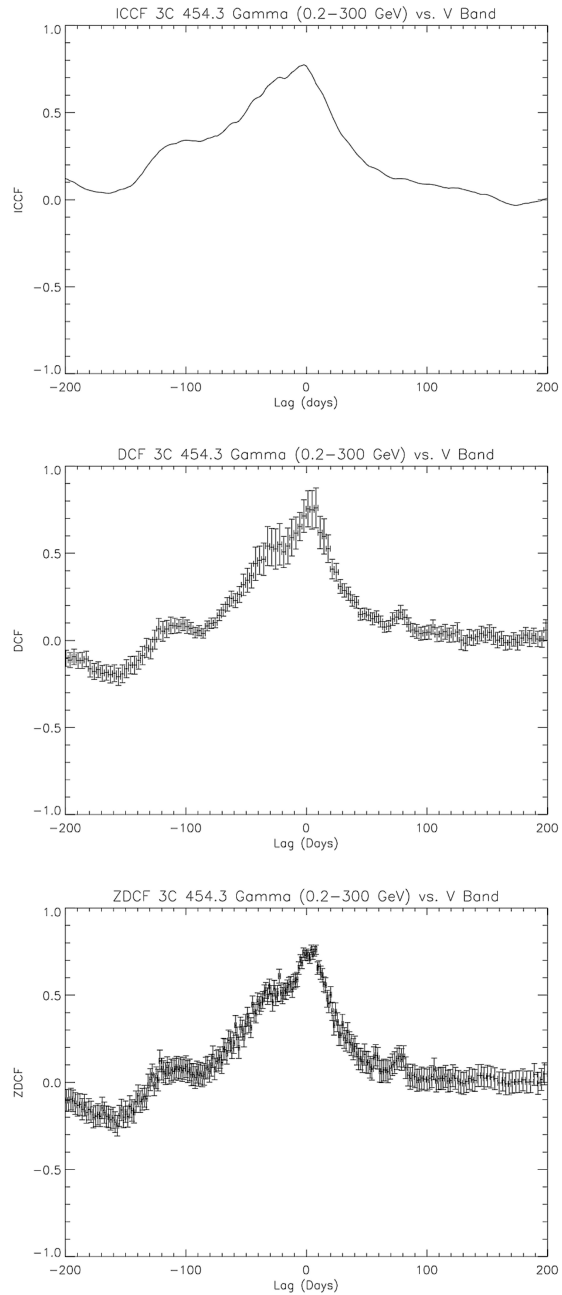}
\caption{Examples of the CCF curves obtained with the three different methods.}
\label{CCF}
\end{figure}

\section{RESULTS}

For the objects in our sample we have light curves in $\gamma$-rays, V band, J, H, K and Ks bands: $\gamma$-rays and NIR light curves for all of them, one object doesn't have a V-band light curve. The lags obtained between the different bands in all objects can be found in Table~\ref{results}. The main conclusions about the sample can be summarized as follows:

\begin{enumerate}
\item We found that in all objects the four NIR bands (J, H, K and Ks) vary simultaneously.
\item No significant difference was found in the results obtained for the OVV, and the results obtained for the BLL.
\end{enumerate}

In the specific case of the source 3C 279, the preliminary analysis of different parts of the available light curves, showed two different significant peaks in the cross-correlation functions obtained with the three methods; this could mean that we have two different delays between the $\gamma$-rays emission and the optical/NIR emissions: One near zero days, and the other in $\sim250$ days.

Our results that in most sources the NIR and optical emissions are well correlated is in agreement with the results obtained by other authors. Our results also confirm the correlations found for PKS 1510-089 and 3C 454.3 between $\gamma$-rays and the optical/NIR emissions (e.g. \citealp{Ma:10a,SMA:12a}).

The multiwavelength analysis will be complemented with millimeter and X-rays data. More details about the analysis will be in a forthcoming paper.

\begin{table}[hd]
\begin{center}
\begin{threeparttable}
\caption{General results of the Cross-Correlation analysis.}
\begin{tabular}{lccc}
\toprule[1pt]
Object & $\gamma$-rays vs. V & $\gamma$-rays vs. NIR & V vs. NIR \\
\midrule[1pt]
3C 66A & N/C & N/C & $11.7^{+7.1}_{-7.0}$ \\
BL Lac & N/C & N/C & $-79\pm13$ \\
Mrk 421 & N/C & N/C & $3.4^{+6.9}_{-6.8}$ \\
Mrk 501 & N/C & N/C & N/C \\
PKS 0716+714 & N/C & N/C & $-1\pm13$$^*$ \\
PKS 2155-304 & $3.5\pm2.3$ & $3.7\pm2.4$ & $-0.7^{+1.6}_{-1.5}$ \\
W Comae & N/C & N/C & $2.1^{+8.3}_{-8.2}$ \\
\hline
3C 273 & N/C & N/C & N/C \\
3C 279 & ** & ** & $0.0^{+3.1}_{-3.0}$ \\
3C 345 & N/C & N/C & $17^{+14}_{-12}$ \\
3C 454.3 & $1.1\pm1.8$ & $-0.4\pm1.8$ & $0.0\pm1.2$ \\
PKS 0235+164 & $22^{+11}_{-8}$ & $18\pm14$ & $1\pm12$ \\
PKS 1510-089 & $7.3^{+4.0}_{-3.3}$$^*$ & $22\pm10$ & $9\pm11$ \\
PKS 1633+382 & N/C & N/C & $2.1^{+8.5}_{-4.5}$ \\
PMN J0808-0751 & N/V & N/C & N/V \\
QSO B0133+476 & N/C & N/C & $-4.5^{+9.5}_{-9.1}$ \\
\bottomrule[1pt]

\end{tabular}
\label{results}
\begin{tablenotes}
\item[*]These results are uncertain, due to problems during analysis.
\item[**]Probable two delays, deeper study needed.
\item N/C: No correlation found. N/V: No V band.
\end{tablenotes}
\end{threeparttable} 
\end{center}
\end{table}%

\bigskip 

\begin{acknowledgments}

This work was supported by CONACyT research grant 151494 (M\'exico). V.P.-A. acknowledges support from the CONACyT program for PhD studies.

\end{acknowledgments}

\bigskip 

\bibliography{references} 

\end{large}
\end{document}